\documentclass[vecphys]{svmult}

\usepackage{makeidx}         % allows index generation
\usepackage{graphicx}        % standard LaTeX graphics tool
\usepackage{multicol}        % used for the two-column index
\usepackage{cite}            % adjusts the "syntax" of the refs in the
\usepackage[bottom]{footmisc}% places footnotes at page bottom
\makeindex             % used for the subject index
\begin{document}

\title{Ground State and Finite Temperature Lanczos Methods}

\author{P. Prelov\v sek$^{1,2}$ and J. Bon\v ca$^{1,2}$}
\institute{J.\ Stefan Institute, SI-1000 Ljubljana, Slovenia
\and Faculty of Mathematics and Physics, University of
Ljubljana, SI-1000 Ljubljana, Slovenia \\
\texttt{peter.prelovsek@ijs.si, }
\texttt{janez.bonca@ijs.si}}

\maketitle

\begin{abstract}

The present review will focus on recent development of exact-diagonali- zation 
(ED) methods that use 
Lanczos algorithm to transform large sparse matrices onto the tridiagonal form. 
We begin with a review  of basic principles  of the 
Lanczos method for computing ground-state static as well as dynamical properties. 
Next, generalization to   finite-temperatures  in the form of well established 
finite-temperature Lanczos method is described. The latter  allows for the evaluation of 
temperatures $T>0$ static and dynamic quantities within various 
correlated models. Several extensions and modification
of the latter method introduced more recently are analysed.  
In particular, the low-temperature Lanczos method and the microcanonical 
Lanczos method, especially  applicable within the high-$T$ regime.
In order to overcome the problems of exponentially growing Hilbert spaces that prevent 
ED calculations  on larger lattices, different approaches based on 
Lanczos diagonalization within the reduced basis  have been developed. In this context, 
recently developed method  based on ED within a limited 
functional space is reviewed.  Finally, we briefly discuss 
the real-time evolution of correlated systems  
far from equilibrium,   which can be simulated using the ED and 
Lanczos-based methods, as well as approaches based on the 
diagonalization in a reduced basis. 

\end{abstract}

\section{Introduction}
\label{sec:lan1}

Models of strongly correlated systems have been one of the most intensively
studied theoretical subjects in the last two decades, stimulated at first by the
discovery of compounds superconducting at high-temperatures and
ever since by the emergence of various novel materials and phenomena
which could be traced back to strongly correlated electrons in these
systems. Recently, cold atoms in optical lattice offer a different
realization of strongly correlated quantum entities, whereby these 
systems can be even tuned closer to theoretical models. 

One of the most straightforward methods to numerically  deal with the 
lattice (discrete) models of correlated particles, which are inherently many-body
(MB) quantum systems, is exact diagonalization (ED) of small-size systems. 
In view of the absence of well - controlled analytical methods,   ED method
has been employed intensively to obtain results for static and dynamical 
properties of various models with different aims: a)  to search and confirm
novel phenomena specific for strongly correlated systems, b)  to test 
theoretical ideas and analytical results, c) to get reference results for 
more advanced numerical techniques. 

MB quantum lattice models of interacting particles are characterized with
a dimension of the Hilbert space given by  
the number of basis states $N_{st} \propto K^N$ that is in turn exponentially increasing 
with the lattice size $N$, where $K$ is the number of local quantum states.
It is therefore clear that ED methods can treat fully only systems
with limited $N_{st}$, i.e., both $K$ and $N$ must be quite modest. 

Among the ED approaches the full ED within the Hilbert space of the model
Hamiltonian, yielding all eigenenergies and 
eigenfunctions,  is the simplest to understand, most transparent and 
easy to implement. In principle it allows the evaluation of any ground state (g.s.) 
property as well as finite temperature $T>0$ static or dynamic quantity, at the expense of 
very restricted $N_{st}$.  In spite of that, it represents a  very instructive approach
but also remains essentially the only practical method when all exact levels
are needed, e.g., for studies of level statistics.

Lanczos-based ED methods have already long history of applications
since Cornelius Lanczos \cite{lanc50} proposed the diagonalization   of
sparse matrices using the iterative procedure, allowing for much bigger
Hilbert spaces $N_{st}$ relative to full ED.  Lanczos diagonalization technique is
at present a part of standard numerical linear algebra procedures 
\cite{parl80,demm97} and as such in solid state physics mainly used to 
obtain the g.s. energies and wavefunction with corresponding expectation 
values. The approach has been quite early on extended to calculation of the 
dynamical response functions within the g.s. \cite{hayd75}. The method has 
been in the last 20 years extensively used in connection with models related to 
high-$T_c$ materials, for which we can refer to an earlier overview 
\cite{dago94}.  

The present review will focus on recent development 
of ED-based and Lanczos-based methods. The basics of the 
Lanczos method are presented in Sec.\ref{sec:lan2} and its 
application for g.s. properties in Sec.\ref{sec:lan3}. 
One of already established 
generalizations is the finite-temperature Lanczos method (FTLM)
\cite{jakl94,jakl00}, reviewed in Sec.\ref{sec:lan4}, which allows for 
the evaluation of $T>0$  static and dynamic properties within 
simplest models. Several
extensions and modifications of the latter have been introduced
more recently, in particular the low-temperature Lanczos method
(LTLM) \cite{aich03} and the microcanonical Lanczos method
(MCLM) \cite{long03}, particularly applicable within the 
high-$T$ regime. 

Since the application of the ED methods there have been attempts
and proposals for the proper choice of reduced basis which could allow
for the study of bigger systems. While this is clearly very broad subject with
most substantial  achievements in one-dimensional (1D) systems within the 
framework of the density-matrix
renormalization-group (DMRG)  idea, there are also 
successful applications in higher D$>1$ combined with the Lanczos
procedure being reviewed in Sec.\ref{sec:lan5}. Recently, there is 
also quite an intensive activity on studies of real-time evolution of 
correlated systems, both under the equilibrium and the 
non-equilibrium conditions 
that  can be simulated using the ED and Lanczos-based methods, 
as discussed in  Sec.\ref{sec:lan6}.  

\section{Exact diagonalization and Lanczos method} 
\label{sec:lan2}

\subsection{Models, geometries and system sizes}
\label{sec:lan21}

ED-based methods are mostly restricted to simple models with
only few local quantum states $K$ per lattice site in order to reach reasonable
system sizes $N$. Consequently, there are only few classes of MB models that 
so far exhaust the majority of ED and Lanczos-method studies,
clearly also motivated and influenced by the  challenging physics and 
relevance to novel materials and related experiments.  

To get some feeling for available sizes 
reachable within the ED-based approaches,  it should be reminded 
that in full ED routines the CPU time scales with the number of operations 
$Op \propto N_{st}^3$, while the 
memory requirement is related to the storage of the whole Hamiltonian
matrix and all eigenvectors, i.e., $Mem \propto N_{st}^2$. This limits at present
stage of computer facilities the full ED method to $N_{st} < 2.10^4$
MB states. On the other hand, using the Lanczos-based iterative methods
for the diagonalization of sparse matrices (Hamiltonians), CPU and 
memory requirements scale as $Op, Mem \propto N_{st}$, at least in their basic  
application, to calculate the g.s. and its wavefunction. In present-day
applications this allows the consideration of much larger 
basis sets, i.e., $N_{st} < 10^9$. Still, lattice sizes $N$ reached using
the Lanczos technique remain  rather modest, compared to some
other numerical approaches as the DMRG and 
quantum-Monte-Carlo (QMC) methods, if the full 
Hilbert basis space relevant for the model is used.  

The simplest nontrivial class of MB lattice models are spin models,
the prototype being the anisotropic Heisenberg model for coupled $S=1/2$
spins,
\begin{equation}
H = \sum_{\langle ij \rangle \alpha} J^{\alpha\alpha} _{ij} S^\alpha_i 
S^\alpha_j, \label{heis}
\end{equation}
where the sum  $\langle ij \rangle$ runs over pairs of lattice 
sites with an arbitrary interaction $J_{ij}^{\alpha\alpha}$ (being in principle 
anisotropic) and $S^\alpha_i$ are component of local $S=1/2$ operator.
The model has $K=2$ quantum states per lattice site and therefore allows for biggest 
$N$ in the ED-based approaches where $N_{st} \propto 2^N$ basis states.
To reduce $N_{st}$ as many symmetries and good quantum numbers
as practically possible are used to
decompose the Hamiltonian into separate blocks. Evident choice are sectors
with the ($z$-component of) total spin $S^z_{tot}$ and the wavevector ${\bf q}$ 
for systems with periodic boundary 
conditions, but further also rotational symmetries of particular lattices
have been used. In this way system sizes up to $N \sim 36$ 
(for largest and most interesting sector $S^z_{tot}=0$) have been reached so 
far using the Lanczos technique without any basis reduction.  

On the basis of this simple model one can already 
discuss the feasibility of the Lanczos-based methods with respect to other
numerical quantum MB methods.  For the g.s. in 1D spin systems 
more powerful methods allowing for much bigger systems are DMRG 
and related approaches. For unfrustrated models in D~$>1$ 
the QMC methods are superior for the evaluation of static quantities
at any $T$. Still, Lanczos-based methods become competitive or at least not 
superseded  for   frustrated spin models (where QMC can run 
into minus-sign problem) or for dynamical properties at $T>0$.  

Next in complexity and very intensively studied prototype model is the 
$t$-$J$ model, representing strongly correlated itinerant electrons
with an antiferromagnetic (AFM) interaction between their spins, 
\begin{equation}
H=- \sum_{\langle ij\rangle  s}( t_{ij} \tilde{c}^\dagger_{js}\tilde{c}_{is}+
{\rm H.c.})+J \sum_{\langle ij\rangle} {\bf S}_i\cdot {\bf S}_j, \label{tj}
\end{equation}
where due to the strong on-site repulsion doubly occupied sites are 
forbidden and one is dealing with projected fermion operators
$\tilde{c}_{is}=c_{is}(1-n_{i,-s})$.  The model can be considered as
a good microscopic model for superconducting cuprates which are doped
Mott insulators. For a theoretical and experimental overview
of Mott insulators and metal-insulator transitions see Ref.\cite{imad98}.
and has been therefore one of the most studied using the Lanczos method \cite{dago94}. 
It has $K=3$ quantum states per lattice site and besides $S^z_{tot}$ and ${\bf q}$, 
also the number of electrons $N_e$ (or more appropriate the 
number of holes $N_h=N-N_e$)
are the simplest quantum numbers  to implement. Since the model 
reveals an interesting physics in D $>$ 1, the effort was in connection
with high-$T_c$ cuprates mostly on 2D square lattice. Here the 
alternative numerical methods have more drawbacks 
(e.g., minus sign problem in QMC methods due to the itinerant character of fermions)
so Lanczos-based methods are still competitive, in particular
for getting information on $T>0$ dynamics and transport.
The largest systems considered with Lanczos method so far are 
2D square lattice with $N=32$ sites and $N_h=4$ holes \cite{leun06}.

Clearly, one of the most investigated within the MB community is 
the standard single-band Hubbard model, which has $K=4$ states
per lattice site. Due to the complexity $N_{st} \propto 4^N$ the
application of ED and Lanczos-based method is already quite restricted 
reaching so far  $N=20$ sites \cite{tohy05} requiring already 
$N_{st} \sim 10^9$ basis states. The model
is also the subject of numerous studies using more powerful
QMC method and various cluster dynamical-mean-field-theory (DMFT)
methods for much larger lattices 
so Lanczos-based approaches have here more specific goals.

\begin{figure}[htb]
\centering
\includegraphics*[width=.9\textwidth]{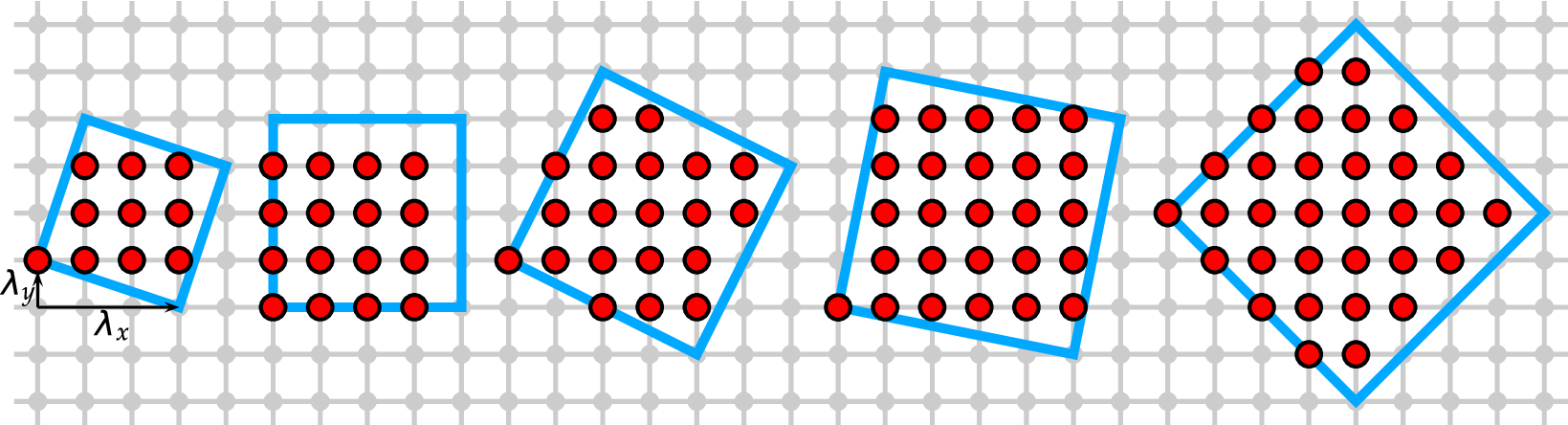}
\caption[]{Tilted clusters used in 2D square-lattice studies}
\label{fig:lanlan1}      
\end{figure}

Since reachable lattices for above mentioned models are rather  small 
it is important  to choose properly their geometries. This is not the 
problem in 1D models, but becomes already essential for 2D
lattices,  analysed in connection with novel materials, in particular
high-$T_c$ cuprates and related materials. In order to keep the periodic
boundary conditions for 2D square lattice the choice of Pythagorean lattices
with $N=\lambda_x^2 + \lambda_y^2$ with $\lambda_x, \lambda_y$ 
\cite{oitm78} has significantly extended available sizes. Some of 
frequently used are presented in Fig.\ref{fig:lanlan1}. Taking into account only even $N$ such
lattices include $N=8, 10, 16, 18, 20, 26, 32, 36$ sites. 
While unit cells of such lattices are squares, it has been observed that
they are not always optimal with respect to the number of 
next-nearest-neighbors and further nearest neighbors. 
It has been claimed and partly tested that better result are obtained
with slightly deformed lattices (still with periodic boundary conditions) 
which at the same time offer an even bigger choice of sizes \cite{bett99}.

\subsection{Lanczos diagonalization technique}
\label{sec:lan22}

The Lanczos technique is a general procedure to transform and reduce a 
symmetric 
$N_{st} \times N_{st}$ matrix $A$ to a symmetric $M \times M$
tridigonal matrix $T_M$. From the chosen initial $N_{st}$-dimensional vector 
${\bf v_1}$ one generates an orthogonal basis of $\{{\bf v_1}, \cdots {\bf v_M} \}$
vectors which span the Krylow space $\{  {\bf v_1},{\bf A v_1}, \cdots
{\bf A^{M-1}} {\bf v_1}\}$ \cite{lanc50,parl80,cull81,demm97}.

In usual applications for the quantum MB system defined with the Hamiltonian operator $H$
the Lanczos algorithm  starts with a normalized vector $|\phi_0\rangle$, chosen as a random 
vector in the relevant  Hilbert space with $N_{st}$ basis states. The procedure
generates orthogonal Lanczos vectors  $L_M = \{|\phi_m\rangle, m=0,M \}$ spanning the Krylow space
 $\{ |\phi_0\rangle, H |\phi_0\rangle \cdots ,H^M |\phi_0\rangle \}$. Steps are as follows:
$H$ is applied to $|\phi_0\rangle$ and the resulting
vector is split in components parallel to $|\phi_0\rangle$, and normalized
$|\phi_1\rangle$ orthogonal to it, respectively,
\begin{equation}
H|\phi_0\rangle=a_0 |\phi_0\rangle + b_1|\phi_1\rangle.
\label{fl1}
\end{equation}
Since $H$ is Hermitian, $a_0=\langle\phi_0|H|\phi_0\rangle$ is real,
while the phase of $|\phi_1\rangle$ can be chosen so that $b_1$ is
also real.  In the next step $H$ is applied to $|\phi_1\rangle$,
\begin{equation}
H|\phi_1\rangle=b_1'|\phi_0\rangle +a_1 |\phi_1\rangle + b_2|\phi_2\rangle,
\label{fl2}
\end{equation}
where $|\phi_2\rangle$ is orthogonal to $|\phi_0\rangle$ and
$|\phi_1\rangle$. It follows also $b_1'=\langle\phi_0|H|\phi_1\rangle
= b_1$. Proceeding with the iteration one gets in $i$ steps
\begin{equation}
H|\phi_i\rangle=b_i|\phi_{i-1}\rangle +a_i |\phi_i\rangle + 
b_{i+1}|\phi_{i+1}\rangle,\qquad 1\leq i \leq M. \label{fl3}
\end{equation}
where in Eq.(\ref{fl3}) by construction there are no terms involving $|\phi_{i-2}\rangle$ 
etc. By stopping the iteration at $i=M$ and setting
$b_{M+1}=0$, the Hamiltonian can be represented in the basis of
orthogonal Lanczos functions $|\phi_i\rangle$ as the tridiagonal
matrix $H_M$ with diagonal elements $a_i, i=0.M$, and
off-diagonal ones $b_i, i=1,M$.  Such a matrix is easily
diagonalized using standard numerical routines to obtain approximate
eigenvalues $\epsilon_j$ and corresponding orthonormal
eigenvectors $|\psi_j\rangle$,
\begin{equation}
|\psi_j\rangle=\sum_{i=0}^M v_{ji} |\phi_i\rangle, \qquad j=0, M.
\label{fl5}
\end{equation}
It is important to realize that $|\psi_j\rangle$ are (in general) not
exact eigenfunctions of $H$, but show a remainder.
On the other hand, it is evident from the diagonalization of $H_M$ that
matrix elements
\begin{equation}
\langle\psi_i|H|\psi_j\rangle=\epsilon_j\delta_{ij}, \qquad i,j=0, M,
\label{fl7}
\end{equation}
are diagonal independently of $L_M$ (but provided $i,j \leq M$), 
although the values $\epsilon_j$ can be only approximate.

If in the equation (\ref{fl3}) $b_{M+1}=0$, we have found a
$(M+1)$-dimensional eigenspace where $H_M$ is already an exact
representation of $H$.  This inevitably happens when $M=N_{st}-1$, but
for $M<N_{st}-1$ it can only occur if the starting vector is
orthogonal to some invariant subspace of $H$ which we avoid by choosing
the input vector $|\phi_0\rangle$ as a random one.

It should be recognized that the Lanczos approach is effective only for sparse
Hamiltonians,  characterized by the connectivity of each basis state with 
$K_n \ll N_{st}$ basis states. All prototype discrete tight-binding models discussed
 in Sec.\ref{sec:lan21} are indeed of such a type in the local MB basis.  Estimating the computation 
requirements, the number of operations $Op$ needed to 
perform $M$ Lanczos iterations scales as $Op \propto K_n MN_{st}$. The main restriction is still 
in memory requirements due to large $N_{st}$. A straightforward application of
Eq.(\ref{fl3}) would require the fast storage of all  $|\phi_i\rangle, i=0,M$, i.e., also the
memory capacity $Mem \propto MN_{st}$. However, for the evaluation of the eigenvalues alone
during the iteration, Eq.(\ref{fl3}), only three $|\phi_i\rangle$ are successively required, 
so this leads to $Mem \propto 3N_{st}$. If the Hamiltonian matrix is not evaluated on the fly
(simultaneously), then also $Mem \propto K_n N_{st}$ for the nonzero Hamilton 
matrix elements is needed.

The Lanczos diagonalization  is in essence an iterative power method
which is known to converge fast for the extreme lower and upper eigenvalues
\cite{parl80,demm97}, clearly in physical application most relevant is the 
search for the g.s. energy $E_0$ and corresponding wavefunction
$|\Psi_0 \rangle$. Typically, $M >50$ are enough to reach very high accuracy for both.
It is evident that for such modest $M \ll N_{st}$ one cannot expect
any reliable results for eigenstates beyond the few at the bottom and the top
of the spectrum. On the other hand, the Lanczos procedure is subject to
roundoff errors, introduced by the finite-precision arithmetics which
usually only becomes severe at larger $M>100$ after the convergence of 
extreme eigenvalues and is seen as the loss
of the orthogonality of vectors $|\phi_i\rangle$.  It can be remedied
by successive reorthogonalization \cite{parl80,cull81,demm97} of new states
$|\phi'_i\rangle$, plagued with errors, with respect to previous
ones. However this procedure requires $Op \sim M^2N_{st}$ operations, and
can become computationally more demanding than Lanczos iterations
themselves. This effect also prevents one to use the Lanczos method, e.g., to
efficiently tridiagonalize full large matrices \cite{demm97}.

\section{Ground State Properties and Dynamics}
\label{sec:lan3}

After $|\Psi_0\rangle$ is obtained, the g.s. static properties can be evaluated
in principle for any operator $A$ as
\begin{equation}
\bar A_0 = \langle \Psi_0 | A | \Psi_0 \rangle.  \label{a0}
\end{equation}
Clearly, the procedure (\ref{a0}) for large basis is effective only if operator $A$ is in the same
basis also sparse, as it is in most cases of interest. 

It is, however, the advantage of the Lanczos procedure that also g.s. dynamical 
functions can be calculated within the same framework \cite{hayd75}. 
Let us consider the  dynamical (autocorrelation) response function 
\begin{equation}
C(\omega) =\langle\Psi_0| A^\dagger \frac{1}{\omega^+ +E_0-H}A|\Psi_0\rangle, \label{fd2}
\end{equation}
for the observable given by the operator $A$  where $\omega^+=\omega+i\epsilon,~\epsilon>0$.  
To calculate $C(\omega)$ one has to run the second Lanczos procedure with
a new initial function $|\tilde\phi_0\rangle$, 
\begin{equation}
|\tilde\phi_0\rangle=\frac{1}{\alpha} A|\Psi_0\rangle, \qquad 
\alpha= \sqrt{\langle\Psi_0| A^\dagger A|\Psi_0\rangle}. \label{fe2}
\end{equation}
Starting with 
 $|\tilde\phi_0\rangle$  one generates another Lanczos subspace $\tilde
L_{\tilde M}=\{|\tilde\phi_j\rangle,\;j=0,\tilde M\}$ with (approximate)
eigenvectors $|\tilde\psi_j\rangle$ and eigenenergies $\tilde \epsilon_j$.
The matrix for $H$ in the new basis is again a tridiagonal one with 
$\tilde a_j$ and $\tilde b_j$ elements,
respectively.  Terminating the Lanczos procedure at given $\tilde M$, one can
evaluate Eq.(\ref{fd2}) as a resolvent of the $H_{\tilde M}$
matrix expressed in the continued-fraction form \cite{mori65,hayd75,dago94},
\begin{equation}
C(\omega)=\frac{\alpha^2}
{\omega^+ +E_0-\tilde a_0-{\displaystyle\frac{\tilde b_1^2}
{\omega^+ +E_0-\tilde a_1-{\displaystyle\frac{\tilde b_2^2}
{\omega^+ +E_0-\tilde a_2-\ldots}}}}}\;, \label{fd3}
\end{equation}
terminating with $\tilde b_{\tilde M+1}=0$, although other termination functions
can also be employed  and well justified.

We note that frequency moments of the spectral function 
\begin{eqnarray}
\mu_l &=&- \frac{1}{\pi} \int_{-\infty}^\infty \omega^l {\rm Im} C(\omega) d\omega
=\langle\Psi_0|A^\dagger(H-E_0)^lA|\Psi_0\rangle= \nonumber \\
&=&\alpha^2  \langle \tilde \phi_0|(H-E_0)^l |\tilde \phi_0\rangle, \label{fd4}
\end{eqnarray}
are exact for given $|\Psi_0\rangle$  provided $l\leq \tilde M$, since 
the operator $H^l, l<\tilde M$, is exactly reproduced within the Lanczos (or corresponding 
Krylow) space $\tilde L_{\tilde M}$.  

Finally, $C(\omega)$ (\ref{fd3}) can be presented as a sum of $j=0,\tilde M$ 
poles at $\omega=\tilde \epsilon_j-E_0$ with corresponding weights $w_j$.
 As a practical matter we note that in analogy to Eq.(\ref{fl5})
\begin{equation}
w_j=|\langle\tilde\psi_j|A|\Psi_0\rangle|^2= \alpha^2 |\langle\tilde\psi_j| \tilde \phi_0\rangle|^2 =
\alpha^2 \tilde v_{j0}^2, \label{wj}
\end{equation}
hence no matrix elements need to be evaluated within this approach.  In
contrast to the autocorrelation function (\ref{fd3}), 
the procedure  allows also the treatment of general correlation
functions $C_{AB}(\omega)$, with $B\ne A^\dagger$.  In this case
matrix elements $\langle\Psi_0|B|\tilde\psi_j\rangle$ have to be
evaluated explicitly.   It should be also mentioned that at least lowest poles of 
$C(\omega)$, Eq.(\ref{fd3}), should coincide with
eigenenergies $\omega=E_i-E_0$ if $| \tilde \phi_0 \rangle$ is not orthogonal
to $| \Psi_0 \rangle$. However, using $\tilde M > 50 $ spurious poles can emerge
(if no reorthogonalization is used) which, however, carry no  weight as
also evident from exact moments (\ref{fd4}).

In this chapter we do not intend to present an overview of applications of the 
full ED and Lanczos-type studies of g.s. static and dynamical properties of 
correlated systems. Such investigations have been numerous even before
the high-$T_c$ era but intensified strongly with studies of prototype models
relevant for high-$T$ cuprates \cite{dago94} and other novel materials
with correlated electrons. Although variety of models have been investigated 
they are still quite restricted in the number of local degrees and sizes.

\section{Static Properties and Dynamics at $T>0$}
\label{sec:lan4}

Before describing the Finite temperature Lanczos method (FTLM)
we should note that the Lanczos basis is very useful and natural
basis to evaluate the matrix elements of the type 
\begin{equation}
W_{kl}=\langle n|H^k B H^l A|n\rangle, \label{fe1}
\end{equation}
where $|n\rangle$ is an arbitrary normalized vector, and $A, B$ are
general operators. One can calculate this expression exactly by
performing two Lanczos procedures with $M=\max(k,l)$ steps. The first
one, starting with the vector $|\phi_0\rangle=|n\rangle$, produces the
Lanczos basis $L_M$ along with approximate eigenstates $|\psi_j\rangle$ 
and $\epsilon_j$. The second
Lanczos procedure is started with the normalized vector $|\tilde \phi_0\rangle
\propto A | \phi_0 \rangle= A | n\rangle$,
Eq.(\ref{fe2}), and generates $\tilde L_M$ with corresponding 
$|\tilde \psi_j\rangle$ and $\tilde \epsilon_j$.
We can now define projectors onto limited subspaces 
\begin{equation}
P_M=\sum_{i=0}^M|\psi_i\rangle\langle\psi_i|,\;\;
\tilde P_M=\sum_{i=0}^M|\tilde\psi_i\rangle\langle\tilde\psi_i|. \label{proj}
\end{equation}
Provided that $(l,k)<M$ projectors $P_M$ and $\tilde P_M$ 
span the whole relevant basis for the
operators $H^k$ and $H^l$, respectively, so that one can rewrite 
$W_{kl}$ in Eq.(\ref{fe1}) as
\begin{equation}
W_{kl}=
\langle \phi_0|P_MHP_MH\ldots HP_MB\tilde P_MH\ldots\tilde P_M
H\tilde P_M A|\phi_0\rangle. \label{fe7}
\end{equation}
Since $H$ is diagonal in the basis $|\psi_j\rangle$ and 
$|\tilde \psi_j\rangle$, respectively, one can write finally
\begin{equation}
W_{kl}=\sum_{i=0}^M\sum_{j=0}^M\langle\phi_0|\psi_{i}\rangle\langle\psi_{i}|
B|\tilde\psi_{j}\rangle\langle\tilde\psi_{j}|A|\phi_0\rangle
(\epsilon_i)^k (\tilde \epsilon_j)^l. \label{fe8}
\end{equation}
It is important to note that matrix element expression (\ref{fe8}) 
is exact,  independent of how (in)accurate representation  $|\psi_i\rangle,
\epsilon_i$ and $|\tilde \psi_j\rangle, \epsilon_j$, respectively, are
to true system eigenvalues. The only condition remains that  number
of Lanczos steps is sufficient, i.e., $M>(l,k)$.

\subsection{Finite temperature Lanczos method: Static quantities}
\label{sec:lan41}

A straightforward calculation of canonical thermodynamic average 
of an operator $A$ at $T>0$ (in a finite system) requires the knowledge 
of all eigenstates $|\Psi_n\rangle$ and corresponding energies $E_n$, 
obtained, e.g., by the full ED of $H$,
\begin{equation}
\langle A\rangle=\sum_{n=1}^{N_{st}}e^{-\beta E_n}\langle \Psi_n|A|
\Psi_n\rangle \biggm/
\sum_{n=1}^{N_{st}} e^{-\beta E_n}, \label{fh1a}
\end{equation}
where $\beta=1/k_BT$. Such direct evaluation is both CPU time and storage 
demanding for larger systems and is at present accessible 
only for $N_{st} \sim 10000$.

In a general orthonormal basis $|n \rangle$ for finite system with $N_{st}$
basis states one can express the canonical expectation value $\langle A\rangle$
as
\begin{equation}
\langle A\rangle=\sum_{n=1}^{N_{st}}\langle n|e^{-\beta H}A|n\rangle
\biggm/
\sum_{n=1}^{N_{st}}\langle n|e^{-\beta H}|n\rangle,
\label{fh1}
\end{equation}
The FTLM for $T>0$ is based on the evaluation
of the expectation value in Eq.(\ref{fh1}) for each starting $|n\rangle$ 
using the Lanczos basis. We note that such procedure guarantees 
correct high-$T$ expansion series (for given finite system) to high order. 
Let us perform the high-$T$ expansion of Eq.(\ref{fh1}),
\begin{eqnarray}
\langle A\rangle &=& Z^{-1}
\sum_{n=1}^{N_{st}}\sum_{k=0}^\infty
\frac{(-\beta)^k}{k!}\langle n|H^kA|n\rangle, \nonumber\\
Z&=&\sum_{n=1}^{N_{st}}\sum_{k=0}^\infty
\frac{(-\beta)^k}{k!}\langle n|H^k|n\rangle. \label{fh2}
\end{eqnarray}
Terms in the expansion $\langle n|H^k A|n\rangle$ can be calculated
exactly using the Lanczos procedure with $M \geq k$ steps (with
$|\phi^n_0\rangle=|n\rangle$ as the starting function)  since this is a
special case of the expression (\ref{fe1}). Using relation
(\ref{fe8}) with $l=0$ and $B=1$, we get
\begin{equation}
\langle n|H^k A|n\rangle=
\sum_{i=0}^M\langle n|\psi^n_{i}\rangle\langle\psi^n_{i}|
A|n\rangle (\epsilon^n_i)^k. \label{fh3}
\end{equation}
Working in a restricted basis $k\leq M$, we can insert the expression
(\ref{fh3}) into sums (\ref{fh2}), extending them to $k >M$.
The final result can be expressed  as
\begin{eqnarray}
\langle A \rangle &=& Z^{-1}\sum_{n=1}^{N_{st}}\sum_{i=0}^M
e^{-\beta \epsilon^n_i}\langle n|\psi^n_i\rangle\langle\psi^n_i|A|n
\rangle, \nonumber \\
Z &=& \sum_{n=1}^{N_{st}}\sum_{i=0}^M e^{-\beta
\epsilon^n_i}\langle n|\psi^n_i\rangle\langle\psi^n_i|n
\rangle, \label{fh4}
\end{eqnarray}
and the error of the approximation is $O(\beta^{M+1})$.

Evidently, within a finite system Eq.(\ref{fh4}), expanded
as a series in $\beta$, reproduces exactly the high-$T$ series to the order
$M$. In addition, in contrast to the usual high-$T$ expansion, Eq.(\ref{fh4})
remains accurate also for $T\to 0$. Let us assume for simplicity that the
g.s. $|\Psi_0\rangle$ is nondegenerate.  For initial states
$|n\rangle$ not orthogonal to $|\Psi_0\rangle$, already at modest
$M\sim 50$ the lowest eigenstate $|\psi^n_0\rangle$ converges to
$|\Psi_0\rangle$. We thus have for $\beta \to \infty$,
\begin{equation}
\langle A\rangle=\sum_{n=1}^{N_{st}}
\langle n|\Psi_0\rangle\langle\Psi_0|A|n\rangle\bigg/
\sum_{n=1}^{N_{st}}\langle n|\Psi_0\rangle\langle\Psi_0|n\rangle 
=\langle\Psi_0|A|\Psi_0\rangle/\langle\Psi_0|\Psi_0\rangle,
\label{fh5}
\end{equation}
where we have taken into account the completeness of the set $|n\rangle$.
Obtained result is just the usual g.s. expectation value of an operator.

The computation of static quantities (\ref{fh4}) still involves the summation 
over the complete set of $N_{st}$ states $|n\rangle$, which is clearly 
not feasible in practice.  To obtain a useful method, a further essential 
approximation replaces the full summation over $| n \rangle$ by a partial one 
over a much smaller set of random states \cite{imad86,silv94}. 
Such an approximation is analogous to
Monte Carlo methods and leads to a statistical error which can be
well estimated and is generally quite small. 
Let us first consider  only the expectation value (\ref{fh1}) with respect
to a single random state $|r\rangle$, which is a linear combination of basis
states
\begin{equation}
|r\rangle=\sum_{n=1}^{N_{st}}\eta_{rn}|n\rangle, \label{fr1}
\end{equation}
where $\eta_{rn}$ are assumed to be distributed  randomly.
Then the random quantity can be expressed as
\begin{eqnarray}
&&\tilde A_r=\langle r|e^{-\beta H}A|r\rangle/
\langle r|e^{-\beta H}|r\rangle =\nonumber \\
&=&
\sum_{n,m=1}^{N_{st}}\eta^*_{rn}\eta_{rm}\langle n|e^{-\beta H}A|m\rangle
\biggm/
\sum_{n,m=1}^{N_{st}}\eta^*_{rn}\eta_{rm}\langle n|e^{-\beta H}|m\rangle.
\label{fr2}
\end{eqnarray}
Assuming that due to the random sign (phase) offdiagonal terms with $\eta^*_{rn} \eta_{rm},
m \neq n$ on average cancel for large $N_{st}$, we remain with  
\begin{equation}
\bar A_r=
\sum_{n=1}^{N_{st}}|\eta_{rn}|^2\langle n|e^{-\beta H}A|n\rangle
\biggm/
\sum_{n=1}^{N_{st}}|\eta_{rn}|^2\langle n|e^{-\beta H}|n\rangle.\label{fr3}
\end{equation}
We can express $|\eta_{rn}|^2=1/N_{st}+\delta_{rn}$. Random
deviations $\delta_{rn}$ should not be correlated with matrix elements
$\langle n|e^{-\beta H}|n\rangle=Z_n$ and $\langle n|e^{-\beta
H}A|n\rangle=Z_n A_n$, therefore $\bar A_r$ is
close to $\langle A\rangle$ with an statistical error 
related to the effective number of terms $\bar Z$ in the
thermodynamic sum, i.e.
\begin{eqnarray}
\bar A_r &=& \langle A\rangle(1 +{\cal O}(1/\sqrt{\bar Z})), \\
\bar Z &=& e^{\beta E_0}\sum_n Z_n = \sum_{n=1}^{N_{st}} 
\langle n| e^{-\beta (H-E_0)} |n \rangle. \label{fr4}
\end{eqnarray}
Note that for $T\to \infty$ we have $\bar Z\to N_{st}$ and therefore
at large $N_{st}$ very accurate average (\ref{fr4}) can be
obtained even from a single random state \cite{imad86,silv94}. On the other
hand, at finite $T<\infty$ the statistical error of $\tilde A_r$
increases with decreasing $\bar Z$. 

To reduce statistical error, in particular at modest $T>0$, within the FTLM 
we sum in addition over $R$ different randomly chosen $|r\rangle$,
so that in the final application Eq.(\ref{fh4}) leads to
\begin{eqnarray}
\langle A \rangle &=& \frac{N_{st}}{ZR}\sum_{r=1}^{R}\sum_{j=0}^M
e^{-\beta \epsilon^r_j}\langle r|\psi^r_j\rangle\langle\psi^r_j|A|
r \rangle, \nonumber \\
Z &=& \frac{N_{st}}{R}\sum_{r=1}^{R}\sum_{j=0}^M e^{-\beta
\epsilon^r_j}|\langle r|\psi^r_j\rangle|^2. \label{fi1}
\end{eqnarray}
Random states $|r\rangle=|\phi^r_0\rangle$ serve as initial functions for 
the Lanczos iteration, resulting in $M$ eigenvalues 
$\epsilon^r_j$ with corresponding $|\psi^r_j\rangle$. 
The relative statistical error is reduced by sampling (both for $\langle A \rangle$ and
$Z$) and behaves as 
\begin{equation}
\delta \langle A\rangle / \langle A\rangle ={\cal O}(1/\sqrt{R\bar Z}).
\label{fr6}
\end{equation}

For general operator $A$ the calculation of $|\psi_j^r\rangle$ and 
corresponding matrix elements
$\langle\psi^r_j|A| r \rangle$ is needed. On the other hand, the
calculation effort is significantly reduced if $A$ is conserved
quantity, i.e., $[H,A]=0$, and can be diagonalized simultaneously
with $H$.  Then
\begin{equation}
\langle A \rangle =
\frac{N_{st}}{ZR}\sum_{r=1}^{R}\sum_{j=0}^M
e^{-\beta \epsilon^r_j}|\langle r|\psi^r_j\rangle|^2 A^r_j. \label{fi2} 
\end{equation}
In this case the evaluation of eigenfunctions is not necessary since
the element $\langle r|\psi^r_j\rangle=v_{j0}^r$, Eq.(\ref{fl5}), 
is obtained directly from eigenvectors of the
tridiagonal matrix $H_M^r$. There are several quantities of interest 
which can be evaluated in this way, in particular the thermodynamic 
properties as internal energy, specific heat, entropy, 
as well as uniform susceptibility  etc. \cite{jakl95,jakl00}.

Taking into account all mentioned assumptions, the
approximation $\langle A \rangle$ (\ref{fi1}) yields a good
estimate of the thermodynamic average at all
$T$. For low $T$ the error is expected to be of the order of ${\cal
O}(1/\sqrt{R})$, while for high $T$ the error is expected to scale
even as ${\cal O}(1/\sqrt{N_{st}R})$. Since arguments leading to these
estimates are not always easy to verify, it
is essential to test the method for particular cases.

\subsection{Finite temperature Lanczos method: Dynamical response}
\label{sec:lan42} 

The essential advantage of the FTLM with respect to other methods is nevertheless 
in the calculation of dynamical quantities.  Let us consider the dynamical susceptibility 
as given by the autocorrelation function $C(\omega)$ (procedure for the general 
correlation function $C_{AB}(\omega)$ is given in Ref.\cite{jakl00}),
\begin{equation}
\chi''(\omega)=\pi(1-e^{-\beta\omega}) C(\omega), \qquad
C(\omega)=\frac{1}{\pi}{\rm Re} \int_{0}^{+\infty} dt  e^{i\omega t} C(t) , \label{chi}
\end{equation}
with
\begin{equation}
C(t)=\langle  A^\dagger(t) A(0) \rangle=
\frac{1}{Z} \sum_n \langle n| e^{(-\beta+it)  H}A^\dagger
e^{-iHt}A |n\rangle. \label{ff1}
\end{equation}
Expanding the exponentials in analogy to static quantities, Eq.(\ref{fh2}), we get
\begin{equation}
C(t) = Z^{-1} \sum_{n=1}^{N_{st}}\sum_{k.l =0}^\infty
\frac{(-\beta+it)^k}{k!}\frac{(-it)^l}{l!}
\langle n|H^k A^\dagger H^lA|n \rangle. \label{ff2}
\end{equation}
Expansion coefficients in Eq.(\ref{ff2}) can be again obtained
via the Lanczos method, as discussed in Sec.\ref{sec:lan41}.  Performing two
Lanczos iterations with $M$ steps, started with normalized
$|\phi^n_0\rangle=|n\rangle$ and $|\tilde\phi^n_0\rangle \propto
A|n\rangle$, respectively, we calculate coefficients $W_{kl}$
following the equation (\ref{fe8}). We again note that (within the full basis 
$|n \rangle$) the series are via $W_{kl}$ exactly evaluated within the 
Lanczos basis up to order $l,k  \leq M$. The latter yields through 
Eq.(\ref{ff2}) a combination of $(\beta,t)$ expansion, i.e. a combination of 
high-$T$, short-$t$ (in frequency high-$\omega$) expansion
to very high order. Extending and resuming series in $k$
and $l$ into exponentials, we get in analogy with Eq.(\ref{fh2})
\begin{equation}
C(t) = Z^{-1} \sum_{n=1}^{N_{st}}
\sum_{i,j=0}^M e^{-\beta \epsilon^n_i} e^{it(\epsilon^n_i-\tilde \epsilon^n_j)}
\langle n|\psi^n_{i}\rangle\langle
\psi^n_{i}|A^\dagger |\tilde\psi^n_{j}\rangle\langle\tilde\psi^n_{j}|A |n\rangle,
\label{ff3}
\end{equation}
Finally replacing the full summation with the random sampling the FTLM
recipe for the correlation function is
\begin{equation}
C(\omega) = \frac{N_{st}}{Z R}\sum_{r=1}^R \sum_{i,j=1}^M
e^{-\beta \epsilon_i} \langle r|\psi_i^r\rangle \langle \psi_i^r | A^\dagger
|\tilde \psi_j^r \rangle \langle \tilde \psi_j^r |r\rangle
\delta(\omega -\tilde \epsilon_j^r +\epsilon_i^r). \label{com}
\end{equation}

We check the nontrivial $T=0$ limit of above expression.  If
$|n\rangle$ are not orthogonal to the g.s. $|\Psi_0\rangle$, then for
large enough $M$ the lowest-lying state converges to $\epsilon^n_0\sim
E_0$ and $|\psi^n_0\rangle\sim|\Psi_0\rangle$, respectively.  In this
case we have 
\begin{equation}
C(\omega,T=0) \approx \frac{N_{st}}{R} \sum_{r=1}^{R}\sum_{j=0}^M
\langle\Psi_0|A^\dagger|\tilde\psi^n_{j}\rangle
\langle\tilde\psi^n_{j} |A| r \rangle \langle r | \Psi_0\rangle  
\delta(\omega + E_0- \tilde \epsilon^n_j) \label{ff4}
\end{equation}
At $T \sim 0$ one needs in general $M\gg 100$ in order that 
at least low-lying states relevant to Eq.(\ref{ff4})  approach 
$|\tilde\psi^n_j\rangle \to |\Psi_j \rangle$ and  $\tilde \epsilon^n_j \to E_j$. 
Also a considerable sampling $R>1$ 
is required to get correct also amplitudes of separate
peaks in the spectrum of  Eq.(\ref{ff4}) which are a subject of a statistical 
error  due to the incomplete projection on different random $|r\rangle$ in 
$\langle\tilde\psi^n_{j} |A| r \rangle \langle r |  \Psi_0\rangle $. Similar statistical
error can in fact appear also for static quantities in Eq.(\ref{fi1}).

\subsection{Finite temperature Lanczos method: Implementation}
\label{sec:lan43}

Most straightforward is the implementation of the FTLM
for static quantities, Eq.(\ref{fi1}). In particular for conserved quantities,
Eq.(\ref{fi2}), the computation load is essentially that of the g.s. Lanczos iteration
repeated $R$ times and only a minor changes are needed
within the usual g.s. Lanczos code.  

On the other hand, for the dynamical
correlation function (\ref{com}) the memory requirement as well as 
the CPU time is dominated mostly by the evaluation of the 
matrix element $\langle \psi_i^r | A^\dagger |\tilde \psi_j^r \rangle$
where the operations scale as $Op \propto R M^2 N_{st}$ 
and memory as $Mem \propto M N_{st}$. 
This effectively limits the application of the FTLM to $50<M<500$ 
where the lower bound is determined by the convergence of the 
g.s. $\vert\Psi_0\rangle $. Still, it should be noted that the calculation
can be done simultaneously (without any additional cost)
for all desired $T$, since matrix elements are evaluated only once.
Evidently, one should use as much as possible symmetries of the
Hamiltonian, e.g., $N_e, S_{tot}^z, {\bf q}$ to reduce effective
$N_{st}$ by splitting the sampling over different symmetry 
sectors.

\begin{figure}[htb]
\centering
\includegraphics*[width=.7\textwidth]{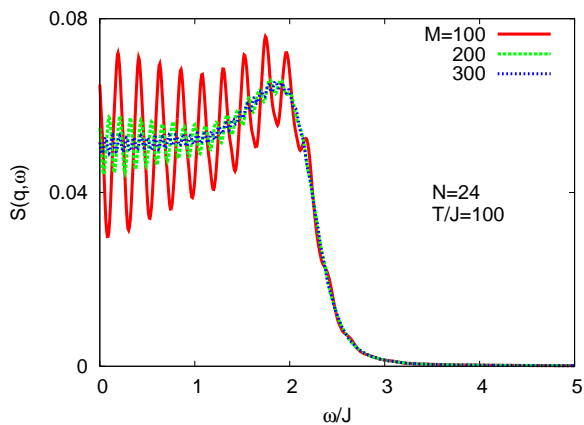}
\caption[]{High-$T$ spin structure factor $S(q=\pi,\omega)$ for the 1D 
Heisenberg model, as calculated with different number of Lanczos steps $M$.}
\label{fig:lan2}   
\end{figure}

The effect of finite $M$ is less evident. Since $M \sim 100$ is
enough to converge well few lowest levels, it is also generally
satisfactory for reliable dynamical correlation functions at low $T$.
At high $T$, however, one can observe very regular oscillations 
which are the artifact of the Lanczos iterations with $M \ll N_{st}$.
Namely, the procedure generates between extreme eigenvalues quite equidistant
spectrum of quasi-states with the level spacing $\Delta \epsilon \sim
\Delta E/M$, where  $\Delta E$ is the full energy span of MB eigenstates.
The effect is well visible in Fig.\ref{fig:lan2} where the high-$T$ result for the 
spin structure factor $S(q=\pi,\omega)$ for the 1D Heisenberg model,
Eq.(\ref{heis}), is presented for various $M$. It is evident that for the
presented case ($N=24$ and $\Delta E \sim 16 J$) 
$M>200$ is sufficient to obtain smooth spectra even
for high $T\gg J$. However, larger $M$ are advisable if sharper structures
persist at high $T$.

The role of random sampling $R$ is less important for intermediate and
high $T$ since the relative error is largely determined via $\bar Z$ as evident
from Eq.(\ref{fr4}). Larger $R \gg 1$ is necessary only for the correct limit 
$T \to 0$ (for given system size) and for off-diagonal operators $A$.

One can claim that the FTLM in general obtains for all reachable systems
results which are at any $T$ very close to exact (full ED) results for the same 
finite (given $N$) system and the accuracy can be improved by 
increasing $M$ and $R$. Still, it remains nontrivial but crucial to understand 
and have in control finite size effects.  

At $T=0$ both static and dynamical quantities
are calculated from the g.s.  $|\Psi_0\rangle$, which can
be quite dependent on the size and on the shape of the system. 
At least in 1D for static quantities the finite-size scaling 
$N \to \infty$  can be performed in a controlled way, although in this case  
more powerful methods as, e.g., the DMRG are mostly available. In higher dimensional 
lattices, e.g., in 2D systems finite-size scaling is practically impossible due to
very restricted choice of small sizes and different shapes.  
Also g.s. ($T=0$) dynamical quantities are often dominated 
by few (typically $N_p<M$) peaks which are finite-size dominated \cite{dago94}.  
On the other hand, $T>0$ generally introduces the thermodynamic averaging 
over a large number of eigenstates. This directly reduces  
finite-size effects for static
quantities, whereas for dynamical quantities spectra become
denser. From Eq.(\ref{com}) it follows that we get in spectra at
elevated $T>0$ typically $N_p \propto RM^2$ different peaks resulting 
in nearly continuous spectra. This is also evident from the example
of a high-$T$ result in Fig.\ref{fig:lan2}.
 
It is plausible that finite-size effects at $T>0$ become weaker. 
However, it should be recognized that there could exist several characteristic 
length scales in the considered physical (and model) system, 
e.g. the antiferromagnetic (AFM) correlation length $\xi$, the transport mean
free path $l_s$ etc.  These lengths generally decrease with increasing $T$
and results for related quantities get a macroscopic relevance provided
that $\xi(T),l_s(T) < L$ where $L \propto N^{1/d}$ is the linear size of the system.
However, there exist also anomalous cases, e.g., in an integrable system $l_s$  can remain 
infinite even at $T \to \infty$ \cite{zoto96,heid07}.  

\begin{figure}[htb]
\centering
\includegraphics*[angle=-90, width=.7\textwidth]{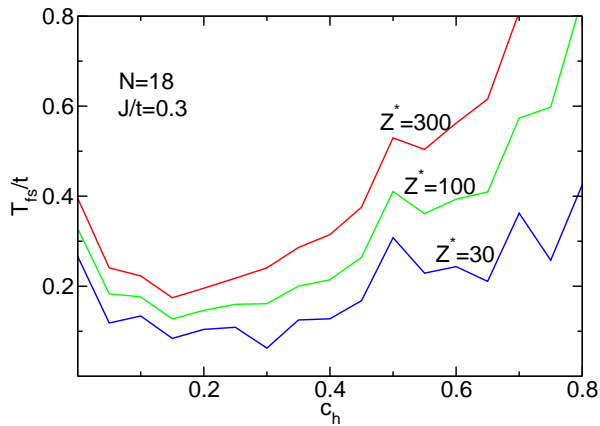}
\caption[]{Finite-size temperature $T_{fs}$ vs. hole doping $c_h$ 
in the 2D $t$-$J$ model with $J/t=0.3$, as calculated with the FTLM 
in system of $N=18$ sites \cite{jakl00}.}
\label{fig:lan3}     
\end{figure}

A simple criterion for finite size effects one can use the normalized
thermodynamic sum $\bar Z(T)$, Eq.(\ref{fr4}), which provides the effective 
number of MB states contributing at chosen $T$ (note that for a system with a
nondegenerate  g.s. $\bar Z(T=0)=1$).  Finite-size temperature $T_{fs}$ can 
be thus defined with the relation $\bar Z(T_{fs}) = Z^*$ where in practice the
range $10<Z^*<50$ is reasonable. Clearly, the FTLM is best suited just 
for systems with a large density of low lying MB states, 
i.e., for large  $\bar Z$ at low $T$.  

Since $\bar Z(T)$ is directly related
to the entropy density $s$ and the specific heat $C_v$ of the system,  
large $\bar Z$ at low $T$  is the signature of 
frustrated quantum MB systems which are generally difficult to cope with
other methods (e.g., the QMC method). Such are typically examples of strongly correlated 
electrons with an inherent frustration, e.g., the doped AFM and the 
$t$-$J$ model, Eq.(\ref{tj}),  in the strong correlation regime $J<t$. 
We present in Fig.\ref{fig:lan3} as an example the variation of $T_{fs}$ 
in the 2D $t$-$J$ model with the hole
doping  $c_h=N_h/N$, as calculated for different $Z^*=30 - 300$ 
for the fixed system of $N=18$ sites  and $J/t=0.3$ as relevant for 
high-$T$ cuprates.   It is indicative that $T_{fs}$ reaches the minimum for 
intermediate (optimum) doping $c_h =c^*_h \sim 0.15$, where we are able to reach
$T_{fs}/t \sim 0.1$. Away from the optimum doping $T_{fs}$ is larger,
i.e., low-energy spectra are quite sparse both for undoped AFM and even more
for effectively noninteracting electrons far away from half-filling (for nearly 
empty or full band).

\subsection{Low Temperature Lanczos Method}
\label{sec:lan44}

The standard FTLM suffers at $T \to 0$ from a statistical error due to finite sampling
$R$, both for the static quantities, Eqs.(\ref{fi1}),(\ref{fr6}), as well as for the dynamical correlations,
Eqs.(\ref{com}),(\ref{ff4}). The discrepancy can be easily monitored by the direct
comparison with the g.s. Lanczos method, Eqs.(\ref{a0},\ref{fd3}). 
To avoid this problem, a variation of the FTLM
method, named Low-temperature Lanczos method  (LTLM) has been proposed
\cite{aich03} which obtains correct g.s. result (for finite systems) independent of 
the sampling $R$.

The idea of LTLM is to rewrite Eq.(\ref{fh1}) in a symmetric form 
\begin{equation}
\langle A\rangle=\frac{1}{Z} \sum_{n=1}^{N_{st}}\langle n|e^{-\beta H/2}Ae^{-\beta H/2} 
|n\rangle, \label{ltlm1}
\end{equation}
and insert the Lanczos basis in analogy with the FTLM, Eq.(\ref{fh1}), now 
represented with a double sum 
\begin{equation}
\langle A \rangle = \frac{N_{st}}{ZR}\sum_{r=1}^{R}\sum_{j,l=0}^M
e^{-\beta (\epsilon^r_j+\epsilon^r_l)/2}  \langle r|\psi^r_j\rangle\langle\psi^r_j|A|
\psi^r_l\rangle \langle \psi^r_l | r \rangle, \label{ltlm2}
\end{equation}
The advantage of the latter form is that it satisfies the correct $T=0$ limit provided
that the g.s. is well converged, i.e., $|\psi^r_0 \rangle \sim   |\Psi_0\rangle$. 
It then follows from Eq.(\ref{ltlm2}),
\begin{equation}
\langle A \rangle = \sum_{r=1}^{R} 
\langle r| \Psi_0\rangle\langle\Psi_0|A| \Psi_0 \rangle \langle \Psi_0|r \rangle
/ \sum_{r=1}^{R} \langle \Psi_0 | r \rangle  
\langle r|\Psi_0\rangle = \langle\Psi_0|A| \Psi_0 \rangle, \label{ltlm3}
\end{equation}
for any chosen set of $|r \rangle$. For the dynamical correlations
$C(t)$ one can in straightforward way derive the corresponding expression
in the Lanczos basis 
\begin{eqnarray}
C(\omega) &=&  \frac{N_{st}}{Z R}\sum_{r=1}^R \sum_{i,j,l=0}^M
e^{-\beta (\epsilon^r_i +\epsilon^r_l)/2} 
\langle r| \psi_i^r\rangle \langle \psi_i^r | A^\dagger
|\tilde \psi_j^{rl} \rangle \langle \tilde \psi_j^{rl} | A | \psi_l^r \rangle 
\langle \psi_l^r |r\rangle \times \nonumber \\
&& \times ~\delta(\omega -\tilde \epsilon_j^{rl} + \frac{1}{2}(\epsilon_i^r+\epsilon_l^r) ). 
\label{ltlm4}
\end{eqnarray}
It is again evident that for $T \to 0$ the sampling does not influence results
being correct even for $R=1$ if the g.s. $|\Psi_0\rangle $ is well converged 
for all starting $|r \rangle$. The payoff is in an additional summation
over the new Lanczos basis which starts from  each $A | \psi_l^r \rangle$ 
in Eq.(\ref{ltlm3}). Since the LTLM is 
designed for lower $T$ there one can effectively restrict summations in $(i,l)$
in Eq.(\ref{ltlm4}) to much smaller $M' \ll M$ where only lowest states
with $\epsilon^r_i ,\epsilon^r_l  \sim E_0$ contribute \cite{aich 03},
and in addition use smaller $M_1 \ll M$ for the basis $|\tilde \psi_j^{rl}\rangle$ .

An  alternative version for Lanczos-type approach \cite{kokado} for
dynamical quantities is not to start the  second Lanczos run from
$A|r\rangle$ \cite{jakl00} or from $A|\psi_l^r\rangle$ \cite{aich03}, but from 
\begin{equation}
  |\widetilde {Ar}\rangle =  \sum_{l=0}^{M}
  A | \psi_l^r\rangle   e^{-\beta \epsilon_l^r /2}
  \langle \psi_l^r |r\rangle. \label{ar}
\end{equation}
In this way one obtains with the second Lanczos run the Lanczos eigenstates
$| \widetilde \psi _k^r\rangle$, which cover the relevant 
Hilbert space for starting random vector $|r\rangle$ 
and the  inverse temperature $\beta$. The resulting dynamical autocorrelation
function is 
\begin{eqnarray}
C(\omega) &=& \frac{N_{\rm st}}{R Z}\sum_{r=1}^R \sum_{i,k=0}^{M}
 e ^{-\beta \epsilon_i^r/2}  \langle r|\psi_i^r\rangle \langle \psi_i^r| A^\dagger | 
 \widetilde \psi_j^r\rangle  \langle \widetilde \psi_j^r |\widetilde {Ar}\rangle 
 \times \nonumber \\
&& \times ~\delta(\omega -\tilde \epsilon_j^r +\epsilon_i^r ). \label{aftlm3}
\end{eqnarray}
In this way the sufficiency of only one random vector in the
$T=0$ limit is reproduced, while at $T>0$ the algorithm
has the same time efficiency as the FTLM, but with much smaller random
sampling needed to reach the same accuracy (at least for low
$T$). However, the price paid is that results for each $T$ need to be
calculated separately, while within the FTLM all $T$
(or $T$ up to certain value within the LTLM) are evaluated simultaneously. 

\subsection{Microcanonical Lanczos Method}
\label{sec:lan45}

While most investigations in strongly correlated systems focus on the 
low-$T$ regime, there are systems where dynamical properties
are nontrivial even at high $T$. Well known such case is the
spin diffusion constant $D_s(T)$ in the isotropic Heisenberg model,
Eq.(\ref{heis}), which is not known by value and moreover not even its 
existence at any $T>0$. Similar although somewhat less controversial 
is the case of transport quantities, both for integrable or generic nonintegrable models.
Whereas the FTLM seems well adapted for studies of transport response
functions, oscillations due to limited $M$ can affect the crucial low-$\omega$
resolution as seen also in Fig.\ref{fig:lan2}. 

At elevated $T$ it is therefore an advantage to use the Microcanonical
Lanczos method (MCLM) \cite{long03}, employing the fact from statistical
physics that in the  thermodynamic limit (for large system)
the microcanonical ensemble   should yield the same results as the 
canonical one. The shortcoming of the MCLM emerges since 
 in finite systems statistical fluctuations
are much larger within the microcanonical ensemble. Still, reachable
finite-size systems have very high density of states in the
core of the MB spectrum as probed by high $T$. Hence, statistical
fluctuations  are at high $T$ effectively smoothed out in contrast to low-$T$
properties dominated by a small number of low lying MB states.

The implementation of the MCLM is quite simple and straightforward.
One first determines the target energy $\lambda = \langle H \rangle(T)$ 
which represents the microcanonical energy equivalent
to the canonical average energy for chosen $T$ and the system size $N$.
Since $\lambda$ is a parameter within the MCLM, one can relate it 
to $T$ by performing either FTLM (simplified due to conserved 
quantity $H$) on the same system or extrapolating full ED results
(with linear dependence  on $N$) on small lattices. Next we 
find a representative microcanonical state $|\Psi_\lambda \rangle$
for the energy $\lambda$. One convenient way within the Lanczos-type
approach is to use the new operator
\begin{equation}
V = (H-\lambda)^2. \label{h2}
\end{equation}
Performing Lanczos iterations with the operator $V$ yields again
the extremum eigenvalues, in particular the lowest one close to
$V \sim 0$. In contrast to the g.s. procedure, the  convergence to a true
eigenstate cannot be reached in system sizes of interest even 
with $M_1 \gg 100$. The reason is extremely small eigenvalue 
spacing of operator $V$ scaling as $\Delta V_n \propto
(\Delta E/N_{st})^2$, $\Delta E$ being the whole energy span within
the given system. Fortunately such a convergence is not
necessary (even not desired) since the essential
parameter is small energy uncertainty $\sigma_E$, given
by
\begin{equation}
\sigma^2_E= \langle \Psi_\lambda | V |\Psi_\lambda \rangle. \label{sige}
\end{equation}
For small energy spread $\sigma_E/\Delta E < 10^{-3}$ typically
$M_1 \sim 1000$ is needed. Again, to avoid storing $M_1$ Lanczos
wavefunctions $|\phi_i\rangle$ Lanczos procedure is performed 
twice as described in Sec.\ref{sec:lan22}, i.e.,  the second time with known 
tridiagonal matrix elements to calculate finally $|\Psi_\lambda \rangle$
in analogy with Eq.(\ref{fl5}).
The latter is then used to evaluate any static expectation average
$\langle A \rangle$ or the dynamical correlation function as
in Eq.(\ref{fd2}),
\begin{equation}
C(\omega,\lambda) =\langle\Psi_\lambda| A^\dagger 
\frac{1}{\omega^+ + \lambda -H} A|\Psi_\lambda \rangle.  \label{ltlmc}
\end{equation}
The latter is evaluated again using Lanczos iterations with 
$M_2$ steps starting with the initial  wavefunction 
$|\tilde \phi_0 \rangle \propto A |  \Psi_\lambda \rangle$ and
$C(\omega)$ is reprepresented in terms of continued fractions. 
Since the MB levels  are very dense and correlation functions 
smooth at $T \gg 0$ large 
$M_2 \gg 100$ are  needed but as well easily reachable to achieve 
high-$\omega$ resolution in $C(\omega)$. 

It is evident that the computer requirement for the MCLM
both regarding the CPU and memory are essentially the same as 
for the g.s. dynamical calculations except that typically 
$M_1, M_2 \gg 100$. In particular requirements are less demanding 
than using the FTLM with $M>100$. A general experience is that for 
systems with large $N_{st} \gg 10000$ the MCLM dynamical 
results agree very well with FTLM results for the same system. 
It should be also noted that actual frequency
resolution $\delta \omega$ in $C(\omega)$, Eq.(\ref{ltlmc}), is limited
by $\delta \omega \sim \sigma_E$ which is, however, straightforward
to improve by increasing $M_1, M_2$ with typical values
$M_1,M_2 >1000$. One can also improve
MCLM results for any $T$ by performing an additional
sampling over initial random starting $|\phi_0 \rangle$
as well as over $\lambda$ with a probability distribution
$p(\lambda)$ simulating the canonical ensemble in a
finite-size system, i.e., by replacing Eq.(\ref{ltlmc}) with
\begin{equation}
C(\omega) = \sum_\lambda p(\lambda) C(\omega,\lambda).
\label{cav}
\end{equation}

\subsection{Statical and dynamical quantities at $T>0$: Applications}
\label{sec:lan46}

The FTLM has been designed to deal with simplest tight-binding
models of strongly correlated electrons, at the time mostly with
challenging microscopic electronic models of high-$T_c$ superconductors 
\cite{jakl94,jakl00}, where besides superconductivity there is a variety of 
anomalous non-Fermi-like properties even in the normal state.
Clearly of interest in this connection are prototype MB
models as the Heisenberg model, Eq.(\ref{heis}), the $t$-$J$ model, Eq.(\ref{tj})
and the Hubbard model on the 2D square lattice. Unfrustrated 
Heisenberg model can be numerically studied on much bigger lattices
with QMC and related methods. The 2D Hubbard model was and still is mostly subject
of DMFT and QMC studies, since at half-filling or close to it
the Lanczos methods are quite restricted due to large $N_{st}$ even
for modest sizes $N \sim 16$. Therefore one focus of Lanczos-based
approaches was on the $t$-$J$ model being with some 
generalizations a microscopic representation of electronic properties
of high-$T_c$ cuprates.

Thermodynamic quantities as chemical potential $\mu$, 
entropy density $s$, specific heat $C_v$ are the 
easiest to implement within the FTLM. Their $T$- and 
(hole) doping $c_h$-dependence within the $t$-$J$ model on a 2D square
lattice (calculated up to $N=26$ sites)  reveal already very anomalous 
behavior of doped Mott insulators \cite{jakl96} (as evident already from 
Fig.\ref{fig:lan3}), confirmed also by results
for more complete Hubbard model \cite{bonc03}. An introduction
of next-neighbor hopping $t'$ generates also an asymmetry in thermodynamic
properties between hole-doped and electron-doped cuprates \cite{tohy03} as well
quite dramatic influence of stripe order  \cite{tohy103} consistent
with the physics of cuprates.

\begin{figure}[htb]
\centering
\includegraphics*[width=.7\textwidth]{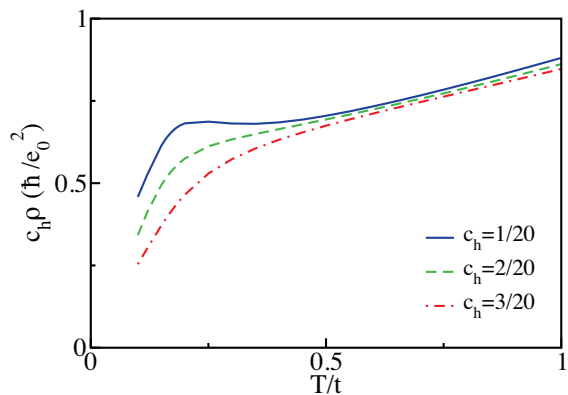}
\caption[]{ Normalized 2D resistivity $c_h \rho$ vs. $T/t$ 
within the $t$-$J$ model with $J/t=0.3$ for different hole concentrations $c_h$ \cite{zeml05}.}
\label{fig:lan4}   
\end{figure}

The advantages of the FTLM and also its feasibility for the
2D $t$-$J$ model are even more evident in quite
numerous studies of spin and charge dynamics at $T>0$ 
\cite{jakl00} which show good agreement with neutron scattering and NMR 
\cite{jakl95,shib01,prel04,bonc04}, optical
conductivity $\sigma(\omega)$ and resistivity $\rho(T)$  \cite{jakl194,zeml05}, 
Hall constant $R_H(T)$ \cite{vebe02} and a general  non-Fermi-liquid behavior 
of cuprates \cite{bonc04}, as well as the puzzling strong influence of
nonmagnetic impurities \cite{prel104}. As an example of a transport quantity
hardly accessible by other methods we present in Fig.\ref{fig:lan4} the 
universal planar resistivity $\rho(T)$, as extracted 
from the dynamical conductivity $\sigma(\omega \to 0)=1/\rho$, 
within the $t$-$J$ model for different doping levels $c_h$ \cite{zeml05}. Result in   
Fig.\ref{fig:lan4}  clearly shows a linear dependence below the pseudogap 
temperature  $T^*$ dependent on  doping $c_h$.
Another characteristic signature is a saturation (plateau) of $\rho(T)$ at low 
doping and the universal trend at high $T$.

\begin{figure}[htb]
\centering
\includegraphics*[width=.8\textwidth]{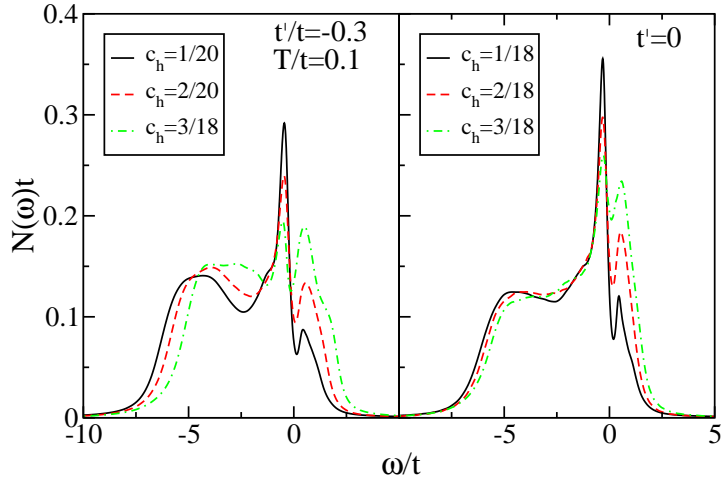}
\caption[]{ Density of states ${\cal N}(\omega)$ for different dopings $c_h$
within the extended $t$-$J$ model with n.n.n. hopping $t'=-0.3t$ and $t'=0$, respectively 
\cite{zeml07}.} \label{fig:lan5}   
\end{figure}

Spectral properties as manifested in single-particle spectral functions 
$A({\bf k}, \omega)$ are in the core of the understanding of cuprates, as well 
as of strongly correlated electrons in general. Here, even g.s. and low-$T$
properties are the challenge for numerical studies whereby the FTLM can
be viewed as a controlled way to get reliable (macroscopic-like) 
$T \to 0$ result, in contrast to quite finite-size plagued results obtained via g.s.
Lanczos procedure \cite{dago94}. Using  the FTLM at $T \sim T_{fs}$ with the
twisted boundary condition can simulate a continuous wavevector 
${\bf k}$. Using in addition the course graining averaging one can reach results for 
$A({\bf k},\omega)$  \cite{zeml07,zeml107, zeml08} giving insight into the
electron vs. hole doped angle-resolved photoemission experiments, 
quasiparticle relaxation and waterfall-like  effects. A characteristic result of such studies is in Fig.\ref{fig:lan5}
for the single-particle density of states ${\cal N}(\omega)= \sum_{\bf k} 
A({\bf k},\omega)$ \cite{zeml07}. Here, the strength of the FTLM is visible 
in the high $\omega$ resolution within the most interesting low-$\omega$ window.
Interesting and reproducible are also nontrivial spectral shapes as the sharp peak close to
$\omega < 0$ and a broad shoulder for $\omega \ll 0$. Most important is, however, 
the evident pseudogap (observed also experimentally in cuprates) visible at 
$\omega \sim 0$ in the low-doping regime.

Besides the challenging models for cuprates there have  been also studies 
of static and dynamical properties of multiband and multiorbital models
which either reduce to the generalized $t$-$J$ model \cite{hors99} or to Kondo
lattice models \cite{hors199,haul00,zere06} and the Falicov-Kimball model \cite{elsh03}. 
While the increasing number of local basis states $K$ clearly limits the applicability
of ED-based methods, they are competitive  in treating nontrivial
frustrated spin models less suitable for the QMC and other methods,  however 
closely related to physics of novel materials. Moreover, frustrated models
are characterized by a large entropy density $s$ and related low $T_{fs}$ essential
conditions for feasibility of FTLM results. Examples of such systems are the
Shastry-Sutherland model \cite{elsh05,elsh07}, 2D $J_1$-$J_2$ model 
\cite{schm07}, and properties of frustrated magnetic molecules \cite{schn10,schn11,
haer11}.

\begin{figure}[htb]
\centering
\includegraphics*[width=.7\textwidth]{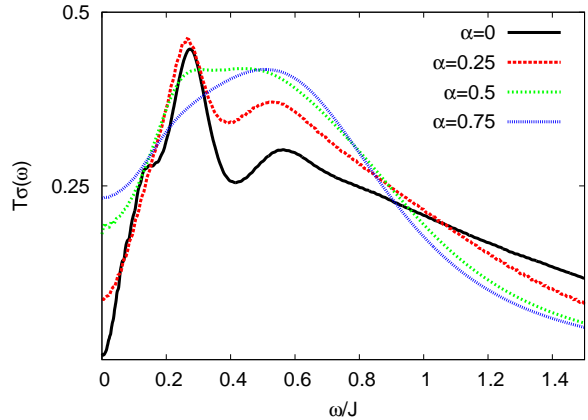}
\caption[]{ High-$T$ dynamical spin conductivity $T \sigma(\omega)$ within the
anisotropic Heisenberg model in the Ising-like regime $\Delta=1.5$ and various 
next-neighbor interaction $\alpha=J_2^{zz}/J$ as calculated with the MCLM
on a chain with $N=30$ sites.}
\label{fig:lan6}   
\end{figure}

Another class of problems which can be quite  effectively dealt with the FTLM 
and MCLM approaches
is the fundamental as well as experimentally relevant problem of transport in
1D systems of interacting fermions as realized, e.g.,  in quasi-1D
spin-chain materials \cite{hess07}. It has been recognized that the transport  response
at any $T>0$ crucially differs between integrable and nonintegrable systems.
Since the 1D isotropic as well as anisotropic Heisenberg model, Eq.(\ref{heis}), 
is integrable it opens a variety of fundamental questions of anomalous transport in 
such systems, the effects of perturbative terms and impurities. Such fundamental questions
on transport and low-$\omega$ dynamic response  remain nontrivial 
even at high $T$ \cite{zoto96,heid07}, hence the MCLM is the most feasible and 
straightforward method. 
It has been in fact first probed on the anomalous transport in 1D insulators \cite{prel204} 
but furtheron used to study interaction-induced transport at $T >0$ in disordered
1D systems \cite{kara09,bari10}, incoherent transport induced by a single either static
\cite{bari09} or dynamical spin impurity \cite{meta10}. 

In Fig.\ref{fig:lan6} we present as an example the MCLM result for the dynamical 
spin conductivity in the anisotropic
Heisenberg model, Eq.(\ref{heis}), where  $J^{zz} \ne J^{xx}=J^{yy}=J $  in the 
Ising-like (with the spin gap in the g.s.) regime $\Delta = J^{zz}/J >1$. 
Results for the high-$T$ dynamical spin conductivity
$T \sigma(\omega)$ are shown for various next-neighbor (anisotropic) coupling
$\alpha = J^{zz}_2/J$. First message is that the MCLM as the method is well adapted for
the high-$\omega$ resolution (here using $M_1=M_2=2000$)
and reaching large $N=30$ ($N_{st} \sim 5.10^6$ in a single $S^z=0,q$ sector).
Another conclusion is that the dynamics of such systems is very anomalous.
For the integrable case $\alpha=0$ we find $\sigma_0= \sigma(0) \sim 0$ 
but also an anomalous finite-size peak at $\omega_p \propto 1/N$ \cite{prel204}. 
At the same time breaking integrability with $\alpha>0$ appears to lead to 
$\sigma_0 >0$ still approaching an 'ideal' insulator (insulating at all $T$) 
for a weak perturbation $\sigma_0(\alpha \to 0) \to 0$ \cite{mier11}.  

\section{Reduced Basis Lanczos Methods}
\label{sec:lan5}

The main shortcoming of ED approaches are finite-size effects that tamper 
calculations on small lattice systems.  Exponentially growing Hilbert spaces 
represent the main obstacle against extending ED calculations to  larger lattices. 
One way to extend ED calculations is to reduce the complete Hilbert space 
and keep only states that give a significant  weight in the g.s. wavefunction or
in the  relevant Hilbert space, e.g., at $T>0$. 
Here, the crucial step represents developing a feasible algorithm for the basis 
reduction. 

One clear example of very effective basis construction is the
DMRG method, most feasible for 1D correlated systems. Within this method, 
described in other chapters, some intermediate steps of ED diagonalization are 
frequently also performed via Lanczos iterations \cite{scho05}. Moreover, there  are
recently developments which are extending the application of DMRG 
procedure to $T>0$ dynamical response \cite{sota08,koka09} involving more directly 
features of Lanczos-based approaches. In particular, recent finite-$T$ dynamical DMRG 
(FTD-DMRG) method \cite{koka09} combines the DMRG selection of MB states 
and the LTLM procedure as described in Sec.\ref{sec:lan43} and in Eqs.(\ref{ar}), 
(\ref{aftlm3}) to  evaluate dynamical correlations at $T>0$, both being effective at low $T$. 
So far the FTD-DMRG method has been applied to find novel features within the 
1D $J_1$-$J_2$ model  \cite{koka10}. 

The ideas to find the reduced basis sets is, however, more general going beyond
the 1D systems and has been successful in solving several problems of
strongly correlated systems.
Even before the discovery of high-$T_c$ superconductors 
that boosted research on models of correlated systems,  Brinkman and 
Rice \cite{brinkman} developed  a string representation of configurational 
space to compute the band renormalization and mobility in the atomic 
limit of the Hubbard model. The string picture was later used to compute  
single hole properties and to estimate hole-pair binding within the $t$-$J$ model 
by many authors \cite{trugman}. In connection with ED calculations on 
finite lattices the truncation scheme based on the limitation of the maximal 
length of strings leads to slow convergence in terms of the number of reduced 
basis states \cite{prel47} in the spin-isotropic limit of the $t$-$J$ model. 
Rather slow convergence is achieved also using  a cluster diagonalization 
scheme together with the systematic expansion of the Hilbert space \cite{riera_dagotto}.

The exact diagonalization method in limited functional space (EDLFS) was originally  
developed specifically for solving the problem of a single charge carrier (hole) doped in the 
AFM  background  \cite{edlfs}. The basic principle of the method is 
based on the above mentioned string picture, that emerges as a moving hole 
through the AFM background creates in its wake paths of overturned spins - strings.
EDLFS was later generalized to include phonon degrees of freedom \cite{tjhol,optic}.
For the purpose of presenting the method we start by considering the 
prototype  $t$-$J$ model, Eq.(\ref{tj}), coupled to dispersionless optical phonons  
on a square lattice
\begin{eqnarray}
H&=&H_t+H_J+H_g+H_{\omega_0} =\nonumber\\
&=& -t\sum_{\langle ij \rangle s} \tilde c^\dagger_{i s}
\tilde c_{ js} + J\sum_{\langle ij\rangle}
\left [ S^z_i S^z_j + \frac{\gamma}{2} (S^+_i S^-_j +S^-_i S^+_j)
\right ] +\nonumber \\
&+&g\sum_i(1-n_i)(a_i^+ + a_i)+ \omega_0 \sum_i a_i^+  a_i, \label{edl_ham}
\end{eqnarray}
where $t$ represents the 
nearest-neighbor hopping, $a_i$ is the phonon annihilation operator,
$n_i=\sum_s n_{is}$ and all double sums run over the nearest neighbor pairs. 
The third term represents electron-phonon
coupling $g$  and the last the energy of
Eistein-type phonons $\omega_0$.
\begin{figure}[htb]
\centering
\includegraphics[width=7cm,clip]{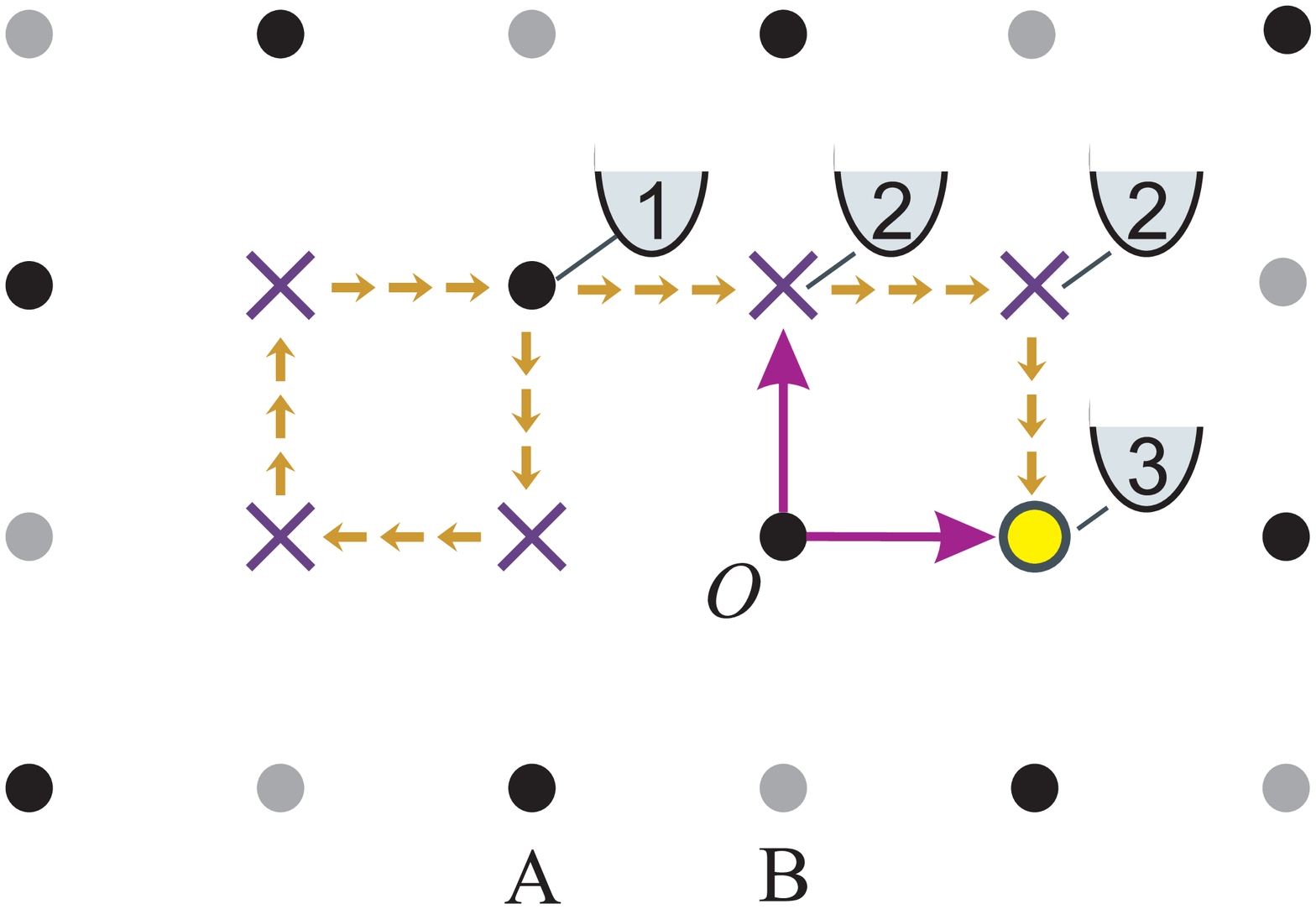}
\caption{(Color online) Schematic representation of a particular  basis
wavefunction obtained using $N_s=10, m=3$, and $N_b=3$. 
Circle represents the hole
position, crosses portray spin flips and numbers indicated excited phonon quanta.
Dots represent lattice sites with no spin flips and zero excited phonon quanta. In
this particular case, $N_{fl}=6$ and $N_p=4$. Presented basis function is one of a
total of $N_{st}=42 \times 10^6$ states, generated using Eq.(\ref{edl_gen}).
}\label{edl_f1}
\end{figure}

The N\' eel state is the g.s. of the model when $t=\gamma=g=0$. 
We construct the limited functional space (LFS) for one hole starting from a
N\'{e}el state on an infinite 2D lattice with a hole placed in the origin  and 
zero phonon degrees of freedom $\vert \phi_0\rangle = c_0 |{\mbox{N\' eel}};0\rangle$. 
We then proceed with the generation of new states, using the following generator
\begin{eqnarray}
\{\vert  \phi_{l}\rangle
\}&=&(H_t+\sum_{i=1}^mH_g^i)^{n_s}\vert \phi_0\rangle,
\label{edl_gen}
\end{eqnarray}
where $H_t$ and $H_g$ are off-diagonal terms in  Eq.\ref{edl_ham}, respectively. 
We select $n_s=1,N_s$  and a fixed value of $m$. 
This procedure generates strings with maximum lengths given by
$N_s$ and phonon quanta at a maximal distance $N_s-1$ from the hole. Parameter $m$
provides the creation of additional phonon quanta leading to a maximum number of phonons
$N_{ph}=m N_s$ in the system. Typically, larger $N_{ph}$ are necessary to achieve convergence in the strong 
electron-phonon  coupling  (polaron) regime. 
While constructing LFS we take into account  explicitly  translational symmetry. 
I.e.,  we store only one out of (in principle)  infinitely many translationally
invariant states, called a parent state.  However, the N\' eel state is a state with a 
broken translational symmetry so allowed translations are
generated by two minimal translations ${\bf r}_{1,2}=(1,\pm 1)$.   
The  basis
wavefunctions are  represented by one of the two nonequivalent hole 
positions ${\bf r}_h$,  sets of strings, representing overturned spins relative 
to the N\' eel state (spin flips),   and
occupation numbers representing excited phonon quanta,
\begin{equation}
\vert  \phi\rangle = \vert {\bf r}_h;{\bf r}_1,{\bf r}_2,\dots,{\bf
r}_{N_{fl}}; n_{{\bf r}_1^\prime},n_{{\bf r}_2^\prime},\dots, n_{{\bf
r}_{N_s}^\prime}\rangle,\label{edl_basis}
\end{equation}
where ${\bf r}_i$ represent spin-flip coordinates, $n_{{\bf r}^\prime_j}$ number of 
phonon quanta at site ${\bf r}^\prime_j$ and
$N_{fl}\leq N_s$ is the total number of spin flips. 

Since the Hilbert space grows exponentially with $N_s$ there exist many 
intuitive physically motivated restrictions optimizing LFS's for different 
parameter regimes  to slow down   the exponential growth. A systematic control over 
the growing Hilbert space can be achieved through realization that 
the diagonal energy of long strings grows linearly with $N_s$. 
Assuming that the contribution of long strings in the g.s. wave-function 
is negligible, we  introduce an additional parameter
$N_b\leq N_s$ that restricts generation of long strings by
imposing a  condition under which   all coordinates of spin flips
satisfy $\vert \mu_h-\mu_f\vert \leq N_b; \mu=\{x,y\}$
where $h$ and $f$ refer to hole and spin-flip indexes,
respectively. Application of this condition improves the quality
of the LFS by increasing the number of states containing
spin flips in the vicinity of the hole while keeping the total
amount of states within  computationally accessible  limits. 

Fig.\ref{edl_f1} represents a particular state generated using Eq.(\ref{edl_gen}). 
Black and grey dots represent sites with spins 'up' and 'down', respectively. 
The hole starts at the position $(-1,1)$  in the direction $(0,-1)$, and travels 
along the  path indicated by arrows. Crosses represent 
spin-flips and numbers represent excited phonon quanta, generated along the 
hole's path. The effect of the parameter $m$ is to set the maximal 
number of excited phonon quanta on indicated positions. 
In this particular parent the hole ends its path  on the B-sublattice. Using allowed 
minimal translations, the position of the hole in the parent state is at 
${\bf r}_h=(1,0)$. 

After the generation of LFS the full Hamiltonian in Eq.(\ref{edl_ham}) is diagonalized 
within this LFS using the standard Lanczos procedure. 
The efficiency of the 
EDLFS method in case of the $t-J$ model  and stability of results against varying 
$N_s$ and $N_b$ has been shown in detail in Ref.\cite{edlfs}.

Besides reliable results 
obtained using the EDLFS there are other important advantages over most other 
methods: a) it  is highly efficient, b) the method is free of finite-size effects, 
c) it treats spin and lattice degrees of freedom on the same footing while 
preserving the full quantum nature of the problem, and d) it allows for 
computation of physical properties at an arbitrary wavevector. Even though 
results depend on the choice of parameters defining the functional generator 
in Eq.(\ref{edl_gen}), such as $N_s$ and $m$, reliable  results are obtained 
already  for relatively small sizes of the Hilbert space $N_{st}$, typically up to
three orders of magnitude smaller than in the case of exact diagonalization 
techniques.  For most static as well as dynamic quantities convergence to 
the thermodynamic limit  can be achieved with a systematic  increase of 
$N_s$ and $m$. 

The EDLFS obviously has few limitations: a) the method is limited to calculations 
in the zero-doping limit, e.g., $N_h=1,2$ mobile particles immersed in an
infinite (AFM) background,  b) the spin anisotropy is inherently built in the method,  and  
c) due to the broken translational symmetry of the starting wavefunction, 
calculations are limited to the reduced AFM Brillouin zone. 

Due to its high efficiency in dealing with spin fluctuation the method represents 
one of the few successful methods that   allows the addition of lattice degrees of
 freedom to a correlated electron model. 
The EDLFS  handles  spin and lattice degrees of
freedom by preserving the full quantum mechanical nature  of the problem  and  
enables a direct calculation of the dynamic response functions in terms of real 
frequency $\omega$. 

The EDLFS can be rather in a straightforward way generalized for the study of the 
two-hole $N_h=2$ problem \cite{bipol,kinet}. 
ED  studies in the $t$-$J$ model for $N_h=2$ were performed  on a 2D square lattices up 
to  $N=32$ sites \cite{leung}. Still, the maximal 
distance between two holes remains  rather small $l_{\mathrm{max}}=\sqrt{N/2}=4$.  
Attempts to increase the lattices sizes beyond ED studies  
led authors to investigate various truncated basis approaches \cite{prel47,riera_dagotto} 
and using small sets of variational wavefunctions with given rotational symmetries 
based on 'string' and 'spin - bag' pictures \cite{wrobel}. 

While the problem of two holes within the $t$-$J$ model represents a challenging problem, the 
addition of quantum phonons seems an almost unachievable 
task. Nevertheless, the EDLFS due to an efficient choice of LFS 
can handle this problem well. 
The construction of the LFS starts from a N\' eel state with holes 
located on neighboring  sites and with zero phonon quanta. Such a state 
represents a parent state of a translationally invariant state. 
We generate new parent states in analogy with Eq.(\ref{edl_gen}) by applying 
the generator of states 
\begin{equation}
\left \{ \vert \phi_l{\rangle}_a\right \} = \left ( H_t + \tilde H_J + 
\sum_{i=1}^mH_{g}^i\right )^{n_s} \vert \phi_{0}{\rangle}_a
\label{edl_genbi}
\end{equation}
where  $\tilde H_J$ denotes the off-diagonal spin exchange term 
 in Eq.(\ref{edl_ham}) applied only to erase spin flips generated through application of $H_t$.
This allows the creation of states with holes positioned further apart that are not 
connected with spin strings. Note that for larger 
$N_s\geq 6$ some of such  states would be generated even without  
evoking $\tilde H_J$ term via the  Trugman loops \cite{trugman}.  
 One of the advantages of this method in comparison
 to other approaches is that it allows much larger distances between the holes, 
 $l_{\mathrm{max}}=N_s+1$.  Note also that the functional generator preserves as well the 
 point-group symmetry. 
 
\begin{figure}[htb]
\centering
\includegraphics[width=0.7\textwidth]{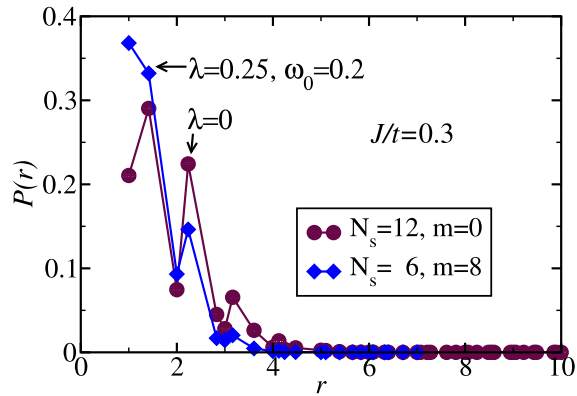}
\caption{
Probability $P(r)$ of finding a hole pair at a distance $r$ within the $t$-$J$-Holstein model
with $J/t=0.3$ and $\omega_0/t=0.2$, evaluated   
for electron-phonon coupling parameters $\lambda=0$ and $\lambda=0.25$. Different 
$N_s$ and $m$ as indicated in legends were chosen giving 
$N_{st} \sim 7.10^6$ and $N_{st} \sim 26.10^6$, respectively.
Both solutions correspond to a $d$-wave symmetry. 
} \label{edl_fig3}
\end{figure}

As an example we present in Fig.\ref{edl_fig3} the probability of finding a 
hole pair at a distance $r$
\begin{equation}
P(r) = \langle \sum_{\langle i \not = j \rangle } n^h_i n^h_j
\delta\left ( \vert {\bf r}_i - {\bf r}_j \vert -r\right )  \rangle,
\label{edl_pr}
\end{equation} 
where $n^h_i=1-n_i$ is local hole density.  
In the case of dimensionless electron-phonon coupling $\lambda=g^2/8 \omega_0 t= 0$, 
the maximal allowed distance between holes 
is $l_{\mathrm{max}}=13$ while for $\lambda=0.25$ $l_{\mathrm{max}}=7$, whereby 
results for $\lambda=0$ are consistent with previous findings using ED
 \cite{leung}, reduced-basis ED \cite{riera_dagotto}, as well as
string picture variational approaches \cite{wrobel}. 

\section{Real Time Dynamics using Lanczos Method} 
\label{sec:lan6}

Research in the field of non-equilibrium dynamics of
complex quantum systems constitutes a formidable theoretical
challenge. 
When dealing with ED approaches or calculations in the reduced basis, 
the time evolution of the time - dependent Shr\" odinger equation,
\begin{equation}
i {\partial \Psi(t) \over \partial t } = H(t)\Psi(t),
\label{time_schrod}
\end{equation}
can be efficiently obtained  using the time - dependent Lanczos technique, 
as originally described in Ref.\cite{park} and later applied and analysed in 
more detail \cite{moha06}. One of the straightforward reasons is 
that most commonly  the Lanczos method is used to compute  g.s.
of MB Hamiltonian. Generalizing the method to time - dependent 
calculation represents only a minor change to already existing codes. Even though 
the method is most suitable for the time evolution of the time - independent 
Hamiltonian, it can nevertheless be applied even to the time - dependent case. 
The time evolution of $\vert \Psi(t)\rangle$ is then
calculated by step-vise change of time $t$  in small time increments $\delta t$, 
generating at each step Lanczos basis of dimension $M$ (typically $M<10$), 
to follow the  evolution  
\begin{equation}
\vert \Psi(t+\delta t)\rangle  \simeq   e^{-i H(t) \delta t} \vert \Psi(t)\rangle
\simeq   \sum_{l=1}^M e^{-i{\epsilon_l} \delta t} \vert{\psi_l\rangle }
{\langle \psi_l\vert} \Psi(t)\rangle ,\label{time_lanc}
\end{equation}
where $\vert \psi_l\rangle, \epsilon_l, l=0,M$ are Lanczos eigenfunctions and eigenvalues, respectively, 
obtained  via the Lanczos iteration started with  $\vert \phi_0\rangle = \vert\Psi(t)\rangle$. 
The advantage of the time-evolution method  following Eq.(\ref{time_lanc}) is that it
preserves the normalization of $|\Psi(t+\delta t)$ for arbitrary large $\delta t$. 
The approximation of finite $M$ in  Eq.(\ref{time_lanc}) is also correct at least  
to the $M$-th Taylor-expansion order in $\delta t$. 
It is, however,  important to stress that $\delta t$ 
should be chosen small enough to take into account properly the 
time-dependence of $H(t)$.
E.g., in the case of driving the system with an external constant 
electric field, $\delta t /t_B\sim 10^{-3}$ where $t_B$ is the Bloch oscillation
 period \cite{marcin_t1, marcin_t2,lev_t1,mier11}. 
  
So far, investigations of correlated driven systems under the influence of a driving
electric field in 1D systems 
using the Lanczos time evolution method
focused on generic systems, like the metallic and Mott - insulating regime within the 1D model of
interacting spinless fermions \cite{marcin_t1,mier11}. Even though rather small systems 
can be studied  it has been established that 
steady state can be reached without  any
additional coupling to a heat bath,  provided that the 
Joule heating of the system is properly taken into account. 

The case of a single charge carrier in an inelastic medium driven  by the external electric 
field in 1D as well as 2D has  been investigated using EDLFS combined with a Lanczos-based
time-evolution  approach \cite{lev_t1,marcin_t1,lev_t2}. 
The strength of EDLFS  is in construction of the Hilbert space that enables not only an 
accurate description of the ground state of the single carrier system but it allows for enough 
extra inelastic (spin or phonon)  excitations to absorb energy, emitted by the field driven carrier, 
until the system reaches the steady state. This again enables a proper description of the 
steady state without coupling the system to an external thermal bath.

\section{Discussion}
\label{sec:lan6}
Exact diagonalization based methods, both the full ED and Lanczos-type ED approach,
  are very extensively employed in the investigations of strongly
correlated MB quantum systems in solid state and elsewhere.  The reason 
for their widespread use are several: a) unbiased approach to the MB
problem without any simplifications or approximations, independent
of complexity of the MB system, 
b) relative simplicity of generating the codes
for various models and observables, c) easy  and straightforward testing
of codes, d) direct interpretation of obtained quantum MB states
and their possible anomalous structure and properties, e) high pedagogical
impact as a quick and at the same time very nontrivial introduction into the
nature of MB quantum physics.  Also the Lanczos-based methods described
in this review, i.e., the g.s. Lanczos method for static and dynamic quantities, 
and somewhat more elaborate FTLM, MCLM, LTLM  and EDLFS, require
rather modest  programming efforts in comparison with more
complex numerical methods,  e.g., the QMC- and
DMRG- based methods, as described in other  chapters. 

Clearly, the main drawback of ED methods is the smallness of lattice
sizes $N$ determined by a limited number of basis states (at present  
$N_{st} < 10^9$) treated with a Lanczos iteration procedure.
The achievable $N$ with ED methods appears quite modest 
in comparison with some established and recently developed 
numerical methods, as the QMC, DMRG, matrix-product-states 
methods etc. Still, in spite of very intensive development 
and advance of novel numerical methods in last two decades
there are several aspects of strong-correlation physics, where ED-based 
methods are so far either the only feasible or at least superior
to other methods. In this chapter we have focused mostly
on Lanczos-based methods and applications where they are
still competitive and get nontrivial results with a macroscopic
validity:  

\noindent a) MB g.s. and its properties of frustrated and complex
models mostly so far do not offer alternative powerful methods except 
ED approaches, at least beyond $D=1$ systems where DMRG-based
methods can be effectively applied,

\noindent b) $T>0$ static properties evaluated with Lanczos-based
methods as the FTLM and the LTLM are as well most powerful and reliable
for frustrated and complex system, in particular in systems with high 
degeneracies of MB states and large entropy at low $T$,

\noindent c) $T>0$ Lanczos methods for dynamical quantities, as the 
FTLM and MCLM, yield for many models and geometries results
superior to other methods or even the only accessible results in several
cases. In particular the advantage of the latter methods is 
high $\omega$ resolution
at all temperatures beyond the finite size limit $T>T_{fs}$,
the macroscopic-like results at low $T$ with a proper scaling 
to  $T \to 0$, and the possibility of detailed studies of systems
with nontrivial (anomalous) dynamics at any, in particular high $T$.

\noindent d) Lanczos technique of ED is also the natural application within 
the methods with a restricted MB basis sets as the EDLFS and
DMRG-type targeting  as well as in the real-time 
evolution studies of strongly correlated systems.

%%%%%%%%%%%%%%%%%%%%%%%%%%%%%%%%%%%%%%%%%%%%%%%%%%%%%%%%%%%%%%%%%%%%%%  }

%%%%%%%%%%%%%%%%%%%%%%%%%%%%%%%%%%%%%%%%%%%%%%%%%%%%%%%%%%%%%%%%%%%%%%

\printindex

\end{document}